\documentclass[twocolumn,prl,showpacs]{revtex4}

\usepackage{graphicx,enumerate}
\usepackage{amsmath}
\usepackage{amssymb}
\usepackage{subfigure}
 \usepackage{verbatim}

\newcommand{\um}{-}
\newcommand{\gl}{\mathrel{\raise0.6ex\hbox{$>$\kern-.75em\lower1ex\hbox{$<$}}}}

\begin{document}

\title{Freezing into Stripe States in Two-Dimensional Ferromagnets and Crossing Probabilities in Critical Percolation}
\author{Kipton Barros}
%\email{kbarros@bu.edu}
\author{P. L. Krapivsky}
\author{S. Redner}
\affiliation{Center for Polymer Studies and Department of Physics, Boston University, Boston, MA 02215}

\begin{abstract}
  When a two-dimensional Ising ferromagnet is quenched from above the
  critical temperature to zero temperature, the system eventually converges
  to either a ground state or an infinitely long-lived metastable stripe
  state.  By applying results from percolation theory, we analytically
  determine the probability to reach the stripe state as a function of the
  aspect ratio and the form of the boundary conditions.  These predictions
  agree with simulation results.  Our approach generally applies to
  coarsening dynamics of non-conserved scalar fields in two dimensions.
\end{abstract}
\maketitle

What is the fate of a kinetic two-dimensional Ising ferromagnet after a
quench from above the critical temperature to zero temperature?  While the
ground state is {\em always} reached in one dimension and {\em never} reached
in three dimensions~\cite{spirin_fate_2001}, the two-dimensional system is
enigmatic.  Previous numerical evidence indicates that a square-lattice
system can either get trapped in an infinitely long-lived metastable stripe
state with probability close to
$\frac{1}{3}$~\cite{spirin_fate_2001,stein_2005,potts}, or reach the ground
state.  In this work, we propose an exact value for the probability for the
two-dimensional system to reach the stripe state, thereby establishing that
the ground state is not necessarily reached.

Our argument is based on relating the {\em non-equilibrium} phenomenon of
coarsening and {\em equilibrium} continuum percolation at the critical point,
We will exploit this unexpected relation to argue that the probability to
reach a stripe state equals $\frac{1}{2}-\frac{\sqrt{3}}{2 \pi} \ln
\frac{27}{16} = 0.3558\ldots$ for free boundary conditions; the corresponding
freezing probability for periodic boundary conditions is 0.3388\ldots.  This
result applies to {\em any} curvature-driven coarsening process with a
non-conserved scalar order parameter, such as the time-dependent
Ginzburg-Landau equation~\cite{lif_62,gunton_dynamics_1983, bray_review}.

\smallskip Our approach is based on two key observations:

\smallskip\noindent (i) Soon after the quench, an emergent characteristic
domain length scale $\ell$ becomes substantially larger than the lattice
spacing $a$, while remaining much less than the system size $L$: $a\ll
\ell\ll L$.  In this regime, this domain mosaic is a realization of the {\em
  critical point of continuum percolation\/}, as previously observed in
various two-dimensional systems~\cite{zallen,arenzon_exact_2007}, and as
argued below.

\smallskip\noindent (ii) In the coarsening regime, $\ell(t)\gg a$, the
dynamics becomes deterministic and domain wall evolution is driven only by
local curvature~\cite{lif_62,gunton_dynamics_1983,bray_review}.  Thus the
global domain topology does not change once the coarsening regime is
reached~\cite{gross_1997}.

\begin{figure*}[ht]
\subfigure[]{\includegraphics[scale=0.11]{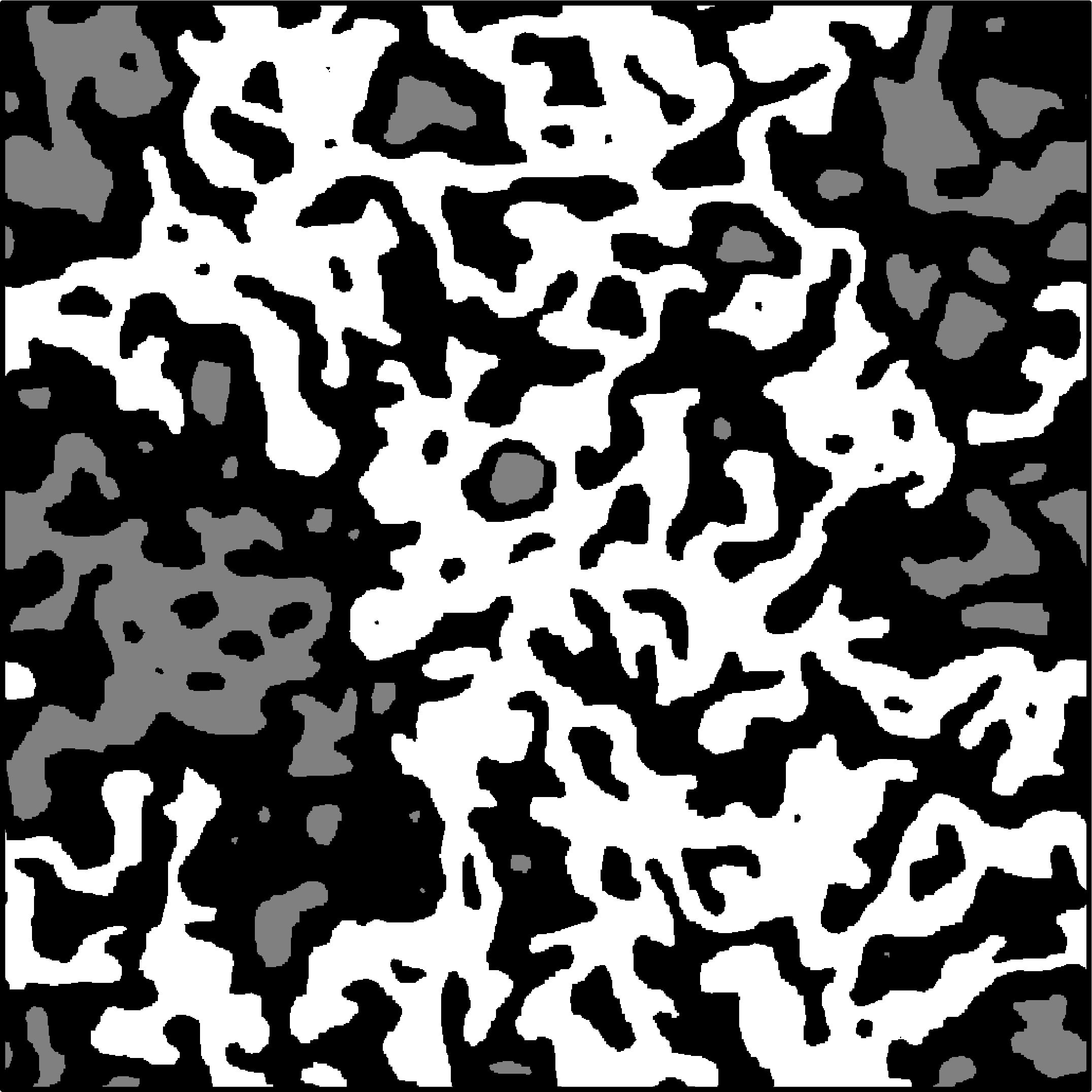}} \quad
\subfigure[]{\includegraphics[scale=0.11]{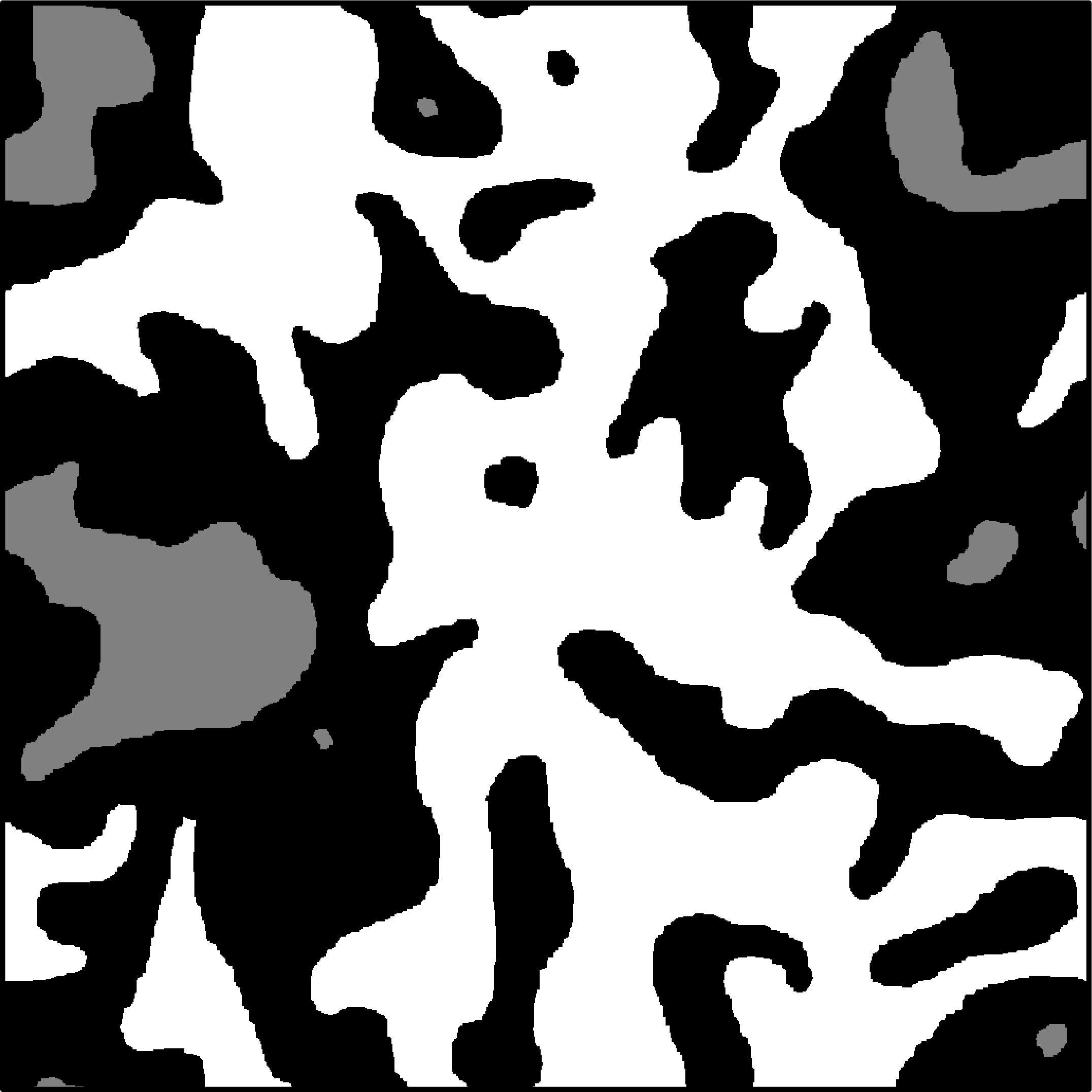}} \quad
\subfigure[]{\includegraphics[scale=0.11]{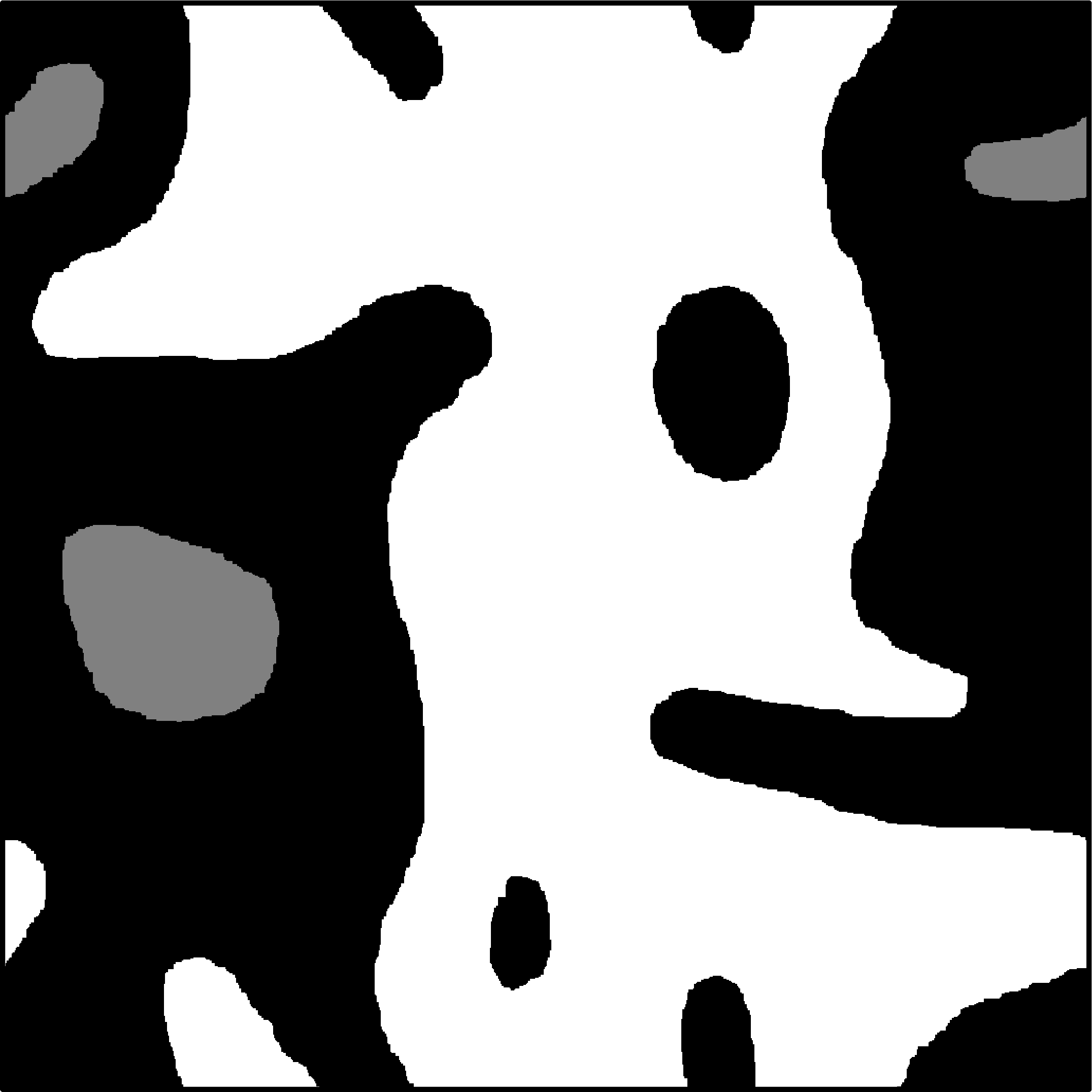}} \quad
\subfigure[]{\includegraphics[scale=0.11]{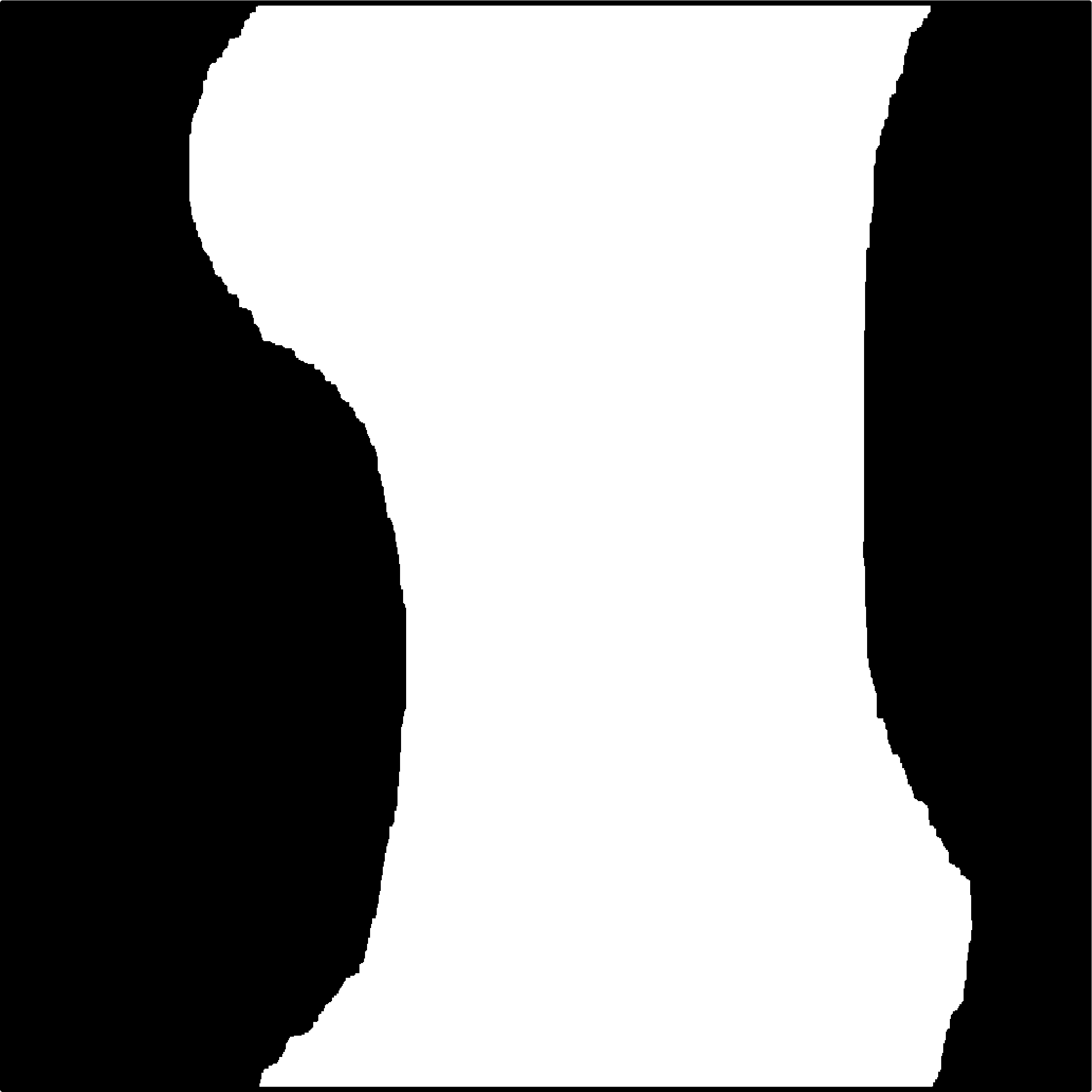}}
\caption{\label{coarsening} Coarsening in the kinetic Ising model on a $1024
  \times 1024$ square lattice with periodic boundary conditions at times (a)
  200, (b) 1000, (c) 5000, and (d) $2.5 \times 10^4$ Monte Carlo steps
  following a quench from $T = \infty$ to $0$.  Domains are regions of either
  spin up (gray) or spin down (black).  A spanning spin-up domain, that
  eventually coarsens into a vertical stripe, is highlighted. }
\end{figure*}

These two features imply that the ultimate fate of the system is
predetermined once the critical percolation state is reached.  For instance,
if a domain exists that crosses the system only horizontally or only
vertically, a stripe state is {\em necessarily} reached.
Figure~\ref{coarsening} shows such an example of a vertical spanning domain
shortly after the quench that ultimately coarsens into a vertical stripe;
conversely, a domain mosaic that spans in both horizontally and vertically
coarsens into a ground state.

\begin{figure}[ht]
\subfigure[]{\includegraphics[scale=0.50]{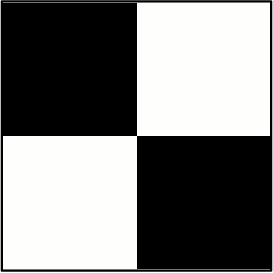}} \quad \quad
\subfigure[]{\includegraphics[scale=0.50]{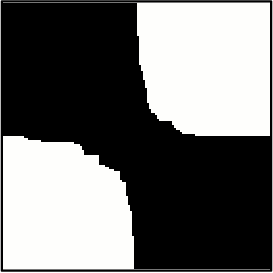}}
\caption{\label{domain-boundary} (a) A state where 4 Ising domains
  meet at a single point is dynamically unstable and evolves into (b)
  a state with no singularities on the domain boundaries.}
\end{figure}

The connection (i) with critical percolation may seem surprising since the
initial fraction of spins of a given sign approaches $\frac{1}{2}$ in the
large-size limit.  This value is below the random site percolation threshold
$p_c\approx 0.5927$ on the square lattice.  However, after a quench to zero
temperature, the spin system quickly approaches the critical state of {\em
  continuum} percolation~\cite{zallen}.  To appreciate this connection, note
that in the coarsening regime (at least the first two panels of Fig.~1), the
concentrations of up and down spins remain very close to $\frac{1}{2}$ and
the boundaries between spin-up and spin-down domains are smooth and do not
contain singular points where four domains meet (as in
Fig.~\ref{domain-boundary}).  This topology coincides with continuum
percolation at its threshold; for example, by a surface whose height
$\phi(x,y)$ is a random function of $(x,y)$ that is symmetric about $\phi=0$.
The regions with positive and negative $\phi$ correspond to the spin-up and
spin-down domains in coarsening, respectively.

While individually observations (i) and (ii) are known, their combined use
allows us to apply exact results about crossing probabilities in percolation
to determine the final state of the kinetic Ising system in two dimensions.
These quantities are defined as the probabilities for the existence of a
spanning cluster with a specified topology.  We denote these crossing
probabilities for free and periodic boundary conditions as $\mathcal{F}$ and
$\mathcal{P}$, respectively.  For a critical rectangular system with aspect
ratio $r$ (ratio of height to width) \cite{aspect}, the crossing
probabilities are non-trivial ({\it i.e.}, strictly between 0 and 1), {\em
  universal\/} functions of $r$ \cite{langlands_universality_1992,
  univ_1996}.  Beautiful exact expressions for various crossing probabilities
were originally calculated via conformal field
theory~\cite{cardy_critical_1992,pinson_critical_1994,watts_crossing_1996,simmons_percolation_2007},
and some of them have been proved in Refs.~\cite{smirnov,schramm,dubedat}.

We begin with the analytically simpler case of free boundary conditions.  In
this setting, every domain mosaic either spans only vertically, only
horizontally, both horizontally and vertically (dual spanning), or the mosaic
does not contain a spanning cluster.  Their respective probabilities,
$\mathcal{F}_{\overline{h}v}$, $\mathcal{F}_{h\overline{v}}$, $\mathcal{F}_{h
  v}$, and $\mathcal{F}_{h v}$ (the dual-spanning and non-spanning
probabilities are identical by up/down symmetry) therefore satisfy the
normalization condition
\begin{equation}
  \label{completeopen} 
\mathcal{F}_{\overline{h}v} +
  \mathcal{F}_{h\overline{v}} 
+  2 \mathcal{F}_{h v}= 1.
\end{equation}
Moreover, the exact form of $\mathcal{F}_{\overline{h}v}$ is known
to be~\cite{watts_crossing_1996, simmons_percolation_2007}
\begin{equation}
  \label{vertopen} 
\mathcal{F}_{\overline{h}v} (r) = \frac{\sqrt{3}}{2 \pi}\,  \lambda\,
  \,\, \text{}_3 F_2 \left( 1, 1, \frac{4}{3} ; \frac{5}{3}, 2 ; \lambda
  \right),
\end{equation}
where $\text{}_p F_q (a_1, \ldots a_p ; b_1, \ldots_{} b_q ; \lambda)$ is the
generalized hypergeometric function~\cite{AS}, $\lambda=\lambda(r)$ is
defined implicitly by
\begin{eqnarray*}
  \lambda  =  \left( \frac{1 - k}{1 + k} \right)^2~,\qquad \mathrm{with}\quad
  r  =  \frac{2 K (k^2)}{K (1 - k^2)}~,\\
\end{eqnarray*}
and $K(u)$ is the complete elliptic integral of the first kind~\cite{AS}.

The corresponding horizontal crossing probability follows by symmetry
\begin{equation}
  \label{horzopen} 
\mathcal{F}_{h\overline{v}}(1/r) = \mathcal{F}_{\overline{h}v}(r).
\end{equation}
while the crossing probability for dual spanning satisfies
\begin{equation}
  \label{dual} \mathcal{F}_{hv}(r) = \frac{1}{2} (1 - \mathcal{F}_{\overline{h}v}(r) - \mathcal{F}_{h\overline{v}}(r)) = \mathcal{F}_{hv}(1/r).
\end{equation}
The basic relation between Ising domains and critical percolation
implies that, {\it e.g.}, the crossing probability
$\mathcal{F}_{\overline{h}v}$ coincides with the probability for the
Ising system to freeze into a vertical stripe state as a function of
$r$.

To test this basic prediction for the freezing probability, we simulate the
kinetic Ising model at zero temperature using single-spin flip dynamics with
the Metropolis acceptance criterion --- a spin is flipped if its energy
decreases or remains the same as a result of the flip.  To make this
simulation more efficient, a list of ``active'' spins --- those whose energy
will not increase upon being flipped --- is maintained and constantly updated
during the dynamics.  In each update step an active spin is chosen at random
and flipped.  One Monte Carlo step corresponds to each active spin flipping
once, on average.

We simulate many quenches from $T = \infty$ to $0$ on lattices of dimension
$(256/r) \times 256$.  For each value of $r$, we performed $2\times 10^4$
simulation runs, with each starting from a different random initial
condition.  We define a domain as a connected cluster of nearest-neighbor
aligned spins.  Clusters are identified using a cluster multilabeling method
\cite{newman_efficient_2000}.  To determine whether a quenched system
ultimately freezes into a stripe state, one should, in principle, simulate
until the system ceases to evolve.  The final stages of the evolution take a
disproportionately large amount of CPU time, however, and it is advantageous
to stop the simulation when the domain mosaic first reaches its final state
topology.  For this Ising system, our simulations indicate that after 200
Monte Carlo steps, domain mosaics have reached the topology of the final
state with a probability that exceeds $0.998$.  Thus we may identify the
state of the system at this early time as the predictor of the topology in
the final state.

\begin{figure}
  \includegraphics[scale=0.87]{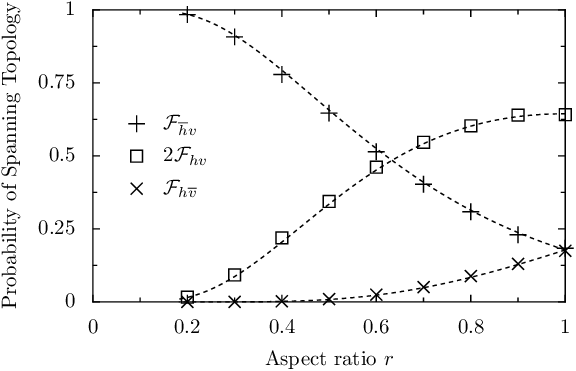}
  \caption{\label{opentop} Probabilities of various domain topologies in the
    kinetic Ising model following a quench with free boundary conditions:
    vertical stripes ($+$), dual-spanning configurations ($\square$), and
    horizontal stripes ($\times$).  Error bars are about $1/3$ the symbol
    size.  The lattice dimensions are $(256/r) \times 256$ for various aspect
    ratios between 0 and 1.  Also shown are the corresponding exact
    percolation crossing probabilities $\mathcal{F}_{\overline{h}v}$,
    $\mathcal{F}_{h\overline{v}}$, and $2 \mathcal{F}_{hv}$, respectively,
    from Eqs.~\eqref{completeopen}--\eqref{dual}. }
\end{figure}

Figure~\ref{opentop} shows our simulation results for the probability
for a specified ultimate fate of an Ising system with free boundary
conditions for a variety of aspect ratios $r$.  The measurements,
labeled using the notation $\mathcal{\hat F}_{(\ldots)}$, agree well
with the corresponding exact crossing probabilities given in
Eqs.~\eqref{vertopen}--\eqref{dual}.  For the important special case
of a square geometry, $r=1$, Eq.~\eqref{vertopen} can be simplified
to~\cite{maier_2003},
\begin{equation}
  \mathcal{F}_{\overline{h}v}(1) = \mathcal{F}_{h\overline{v}}(1) = \frac{1}{4} -
  \frac{\sqrt{3}}{4 \pi} \ln \frac{27}{16} = 0.1779\ldots
\end{equation}
from which the probability of the Ising system coarsening into a stripe state
equals $2 \mathcal{F}_{\overline{h}v}(1)=0.3558\ldots$.  An earlier numerical
estimate for the probability of reaching a stripe
state~\cite{spirin_fate_2001} is consistent with this exact result.

For periodic boundary conditions, a parallel set of results can be
constructed to again connect the ultimate fate of the Ising system and
percolation crossing probabilities.  The nature of the crossing probabilities
is substantially more complex for systems with periodic boundaries because
spanning clusters can wrap around the torus multiple times in the vertical
and horizontal directions.  There are two types of spanning
clusters~\cite{pinson_critical_1994,pruessner_winding_2004,F}.  ``Winding''
clusters are labeled by their vertical and horizontal winding numbers,
$(a,b)$.  For example, winding numbers $(0,1)$ and $(1,0)$ correspond to a
vertical and a horizontal stripe, respectively. A spanning cluster that wraps
around the torus once in the vertical direction and once in the horizontal
direction can have one of two winding number pairs, $(1,1)$ or $(1,-1)$, and
gives a diagonal stripe configuration when the torus is unrolled onto the
square.  The other cluster type is the ``cross topology'' in which a spanning
cluster is formed by the union of two or more spanning clusters with distinct
winding numbers.

Let $\mathcal{P}_{a,b}(r)$ denote the crossing probability for a spanning
cluster with winding numbers $(a,b)$ to exist on a rectangle with aspect
ratio $r$ and periodic boundary conditions.  This probability is given
by~\cite{pinson_critical_1994},
\begin{eqnarray}
\label{windingclusters}
  \mathcal{P}_{a, b} (r)& \!\!=\!\! & \sum_{l \in \mathbb{Z}} \left[ Z_{3al,3bl}
   \!-\!  Z_{2al,2bl}
\!-\!\frac{1}{2} Z_{(3l\!+\!1)a,(3l\!+\!1)b}\right. \nonumber\\
&&~~~~~\!-\!\left. \frac{1}{2} Z_{(3l\!+\!2)a,(3l\!+\!2)b}
   \!+\!   Z_{(2l\!+\!1)a,(2l\!+\!1)b}\right]\!,
\end{eqnarray}
where $Z_{m,n}$ is a shorthand for $Z_{m,n}(\frac{2}{3};r)$; generally
\[ Z_{m, n} (g ; r) = \frac{\sqrt{g}}{\sqrt{r} \eta^2 (e^{\um 2 \pi r})}\,\,
e^{-\pi g(m^2/r + n^2r)},\] and $\eta(q) = q^{1/24} \prod_{k \geq 1}(1-q^k)$
is the Dedekind $\eta$ function~\cite{AS}.  Additionally, the configuration
with cross topology occurs with
probability~\cite{pinson_critical_1994}
\begin{equation}
  \label{crosscluster} \mathcal{P}_X (r) = \frac{1}{2}  \left[ Z\!\left(
  \frac{8}{3}, 1 ; r \right) - Z\!\left( \frac{8}{3}, \frac{1}{2} ; r \right)
  \right],
\end{equation}
where $Z(g,f;r) = f \sum_{m, n \in \mathbb{Z}} Z_{fm, fn} \left( g ; r
\right)$.  By symmetry, $\mathcal{P}_X (r)$ also represents the probability
that no cluster spans the system.  

\begin{figure}
  \includegraphics[scale=0.87]{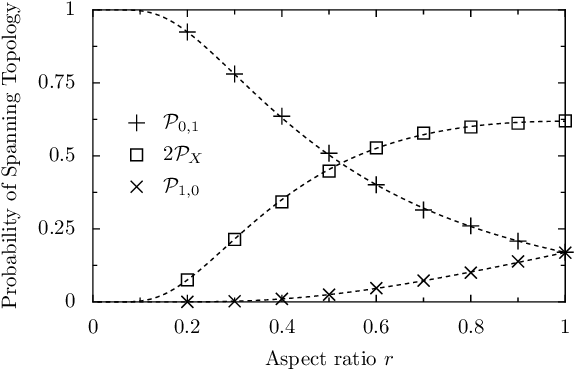}
  \caption{\label{pertop} Probabilities of various domain topologies in the
    kinetic Ising model following a quench with periodic boundary conditions:
    vertical stripes ($+$), dual-spanning configurations ($\square$), and
    horizontal stripes ($\times$), with error bars about $1/3$ of the symbol
    size.  The lattice dimensions are $(256/r) \times 256$ for aspect ratios
    between 0 and 1. Also shown are the corresponding exact percolation
    crossing probabilities $\mathcal{P}_{1,0}$, $\mathcal{P}_{0,1}$, and $2
    \mathcal{P}_{X}$, respectively, from Eqs.~\eqref{windingclusters} and
    \eqref{crosscluster}.  }
\end{figure}

To compute the crossing probabilities numerically, it is preferable to employ
the asymptotic expansions~\cite{pruessner_winding_2004}
\begin{eqnarray*}
  \mathcal{P}_{0, 1} (r) & = & 1 - 2 \rho^{5/4}+\rho^3+2\rho^{12}-4\rho^{53/4}+3\rho^{15}
 \ldots  \\
  \mathcal{P}_{1,0}(r) & = & \sqrt{\frac{2 r}{3}} 
\left(\rho^3-\rho^{15}-\rho^{27}+4\rho^{35}\ldots\right)\\
  \mathcal{P}_X (r) & = & \rho^{5/4} -\rho^3-2\rho^{12}+ 2\rho^{53/4}-2\rho^{15}\ldots,
\end{eqnarray*}
where $\rho\equiv e^{-\pi/6r}$, rather than evaluating the special functions
directly.  These expansions provide an excellent approximation for the entire
range of aspect ratio $0<r<1$~\cite{pruessner_winding_2004}.

Again, a domain mosaic characterized by winding numbers (0,1) [or (1,0)]
occurs with probability $P_{0,1}$ (or $P_{1,0}$), as given by
Eq.~\eqref{windingclusters}, and coarsens into a vertical (or a horizontal)
stripe state.  Similarly, a mosaic with cross topology occurs with
probability $2 \mathcal{P}_X$ given by Eq.~\eqref{crosscluster}, and coarsens
directly into the ground state (either all spins pointing up or all pointing
down).  However, a domain mosaic can also reach the ground state by the
indirect route of first forming a diagonal stripe state with non-zero winding
numbers in both directions.  As found previously for the specific case of the
$(1,1)$ stripe, such states are long lived~\cite{spirin_fate_2001};
namely, they reach the ground state at a time scale that is much larger than
the typical coarsening time $\mathcal{O}(L^2)$.

In Fig.~\ref{pertop} we plot the realizations that evolve to a topology with
winding number $(0,1)$ or $(1,0)$, or to the cross topology for a variety of
aspect ratios $r$.  These again agree well with the exact percolation
crossing probabilities that follow from Eqs.~\eqref{windingclusters} and
\eqref{crosscluster}.  In the specific case of the square system (aspect
ratio $r=1$), the probability of reaching an infinitely long-lived stripe
state is $0.3388\ldots$.  Because the kinetic Ising model can also evolve to
diagonal stripe topologies, $\mathcal{P}_{0, 1} + \mathcal{P}_{1, 0} + 2
\mathcal{P}_X$ is less than 1.

In conclusion, the probabilities with which Ising ferromagnets freeze into
metastable stripe states correspond exactly to crossing probabilities in
critical continuum percolation.  This correspondence relies on the initial
statistical symmetry between up and down spins, which applies when the system
is quenched from equilibrium at {\em any} supercritical initial temperature,
$T>T_c$.  Our simulation results for the probabilities to reach a specified
ultimate fate (Figs.~\ref{opentop} and \ref{pertop}) are in excellent
agreement with theoretical predictions.  Our approach can be applied to
arbitrarily-shaped domains and boundary conditions, and also can be used to
determine more subtle characteristics, such as the distribution of the number
of stripes.

Our theory generally applies to phase ordering kinetics in two dimensions
with non-conserved scalar order parameter, as well as to quenches to non-zero
subcritical temperatures.  In the latter case, the coarsening regime requires
that $\xi \ll \ell(t) \ll L$ (where the equilibrium correlation length $\xi$
may be arbitrarily large).  Metastable stripe states will now persist for a
finite but very long time compared to the coarsening time scale before the
final approach to the equilibrium state.

\begin{acknowledgments}
  \smallskip\noindent We are grateful to R. Ball and W. Klein for very useful
  conversations. This work is supported by DOE grant DEFG-0295ER14498 and NSF
  Grant DGE-0221680 (KB), NSF grant CCF-0829541 (PLK), and NSF grant
  DMR0535503 (SR).
\end{acknowledgments}


\begin{thebibliography}{99}
\bibitem{aspect} We deal only with rectangular systems and the thermodynamic
  limit is taken with a fixed aspect ratio $r$.  Taking the limit in this
  specific way is necessary since the crossing probabilities depend on $r$.
\end{thebibliography}
\end{document}